\documentclass[aps,twocolumn,pra]{revtex4-1}
\usepackage{graphics}
\usepackage{dcolumn}
\usepackage{bm}
\usepackage{amsmath}
\usepackage{hyperref}
\hypersetup{colorlinks=true, urlcolor=blue}
\usepackage{color}
\usepackage{tikz}
\usepackage{multirow}

\usepackage{float}
\usepackage[caption = false]{subfig}
\usepackage{newlfont}
\usepackage{amssymb}
\usepackage{amsfonts}
\usepackage{amsthm}
\usepackage{graphicx}
\usepackage{bm}

\def \be{\begin{equation}}
\def \ee{ \end{equation} }

\begin{document}

\definecolor{red}{rgb}{1,0,0}
\title{On the strong superadditivity of entanglement of formation\\ --Conditions and examples}
\author{Asutosh Kumar}
\email{asutoshk.phys@gmail.com} 

\affiliation{
P. G. Department of Physics, Gaya College, Magadh University, Rampur, Gaya 823001, India\\ Harish-Chandra Research Institute, HBNI, Chhatnag Road, Jhunsi, Prayagraj 211 019, India\\ Vaidic and Modern Physics Research Centre, Bhagal Bhim, Bhinmal, Jalore 343029, India}

\begin{abstract}
From Shor's [Comm. Math. Phys. {\bf 246(3)}, 453 (2004)] and Hastings's [Nature Phys. {\bf 5}, 255 (2009)] studies, the strong superadditivity of entanglement of formation is, in general, not true. 
In this paper we provide conditons for strong superadditivity of entanglement of formation, and consequently identify classes of states that are strongly superadditive. 

\end{abstract}

\maketitle
  
{\it Introduction}.-- 
Entanglement \cite{schrodinger, horodecki09}, a quantum property of quantum systems, is a valuable resource in many quantum information processing tasks including communication, computation and cryptography \cite{nielsen, preskill, wilde}. Characterization and quantification of entanglement, therefore, is necessary. An entanglement measure essentially quantifies {\em how much} a quantum state $\rho_{AB}$ is entangled; $\rho_{AB}$ is entangled if it cannot be written in the separable form
\begin{equation}
\rho_{AB} = \sum_i p_i \rho_A ^i \otimes \rho_B^i.
\end{equation} 

For a pure state $|\psi\rangle_{AB}$ of a quantum system composed of the subsystems A and B, a bona fide measure of entanglement is given by the von Neumann entropy of the reduced density matrix
\begin{equation}
E(|\psi \rangle_{AB}) = S(\mbox{tr}_B |\psi\rangle_{AB} \langle \psi|) = S(\mbox{tr}_A |\psi\rangle_{AB} \langle \psi|),
\label{eof-def1}
\end{equation}
where $S(\rho) = -\mbox{tr} (\rho \ln \rho)$. However, for a mixed state the situation is more evolved, and there are several entanglement measures in literature \cite{horodecki09}.  Entanglement of formation (EoF) \cite{eof}, based on the {\em convex roof} construction \cite{convexroof}, is a widely used measure, defined by
\begin{equation}
E_F^{A|B}(\rho_{AB}) = \mbox{inf} \sum_i p_i E^{A|B}(|\psi_i\rangle_{AB}),
\label{eof-def2}
\end{equation}
where ``infimum'' is taken over all pure state decomposition of $\rho_{AB} = \sum_i p_i |\psi_i\rangle_{AB} \langle \psi_i|$, and superscript $A|B$ denotes the cut across which entanglement is computed. 
We will call an ensemble $\{p_i, |\psi_i \rangle_{AB} \}$ which realizes $\rho_{AB}$, and for which the minimum is attained the {\em optimal ensemble}.
EoF may be interpreted as the minimal pure-states entanglement required to build up the mixed state. It enjoys several appealing properties of an entanglement measure: it vanishes for separable states, is convex and monotone \cite{eof}, is known to be continuous \cite{eof-continuity}, and $E_F(|\psi\rangle_{AB}) = E(|\psi\rangle_{AB})$, by construction. Though it is not easy to evaluate, an explicit analytical expression of EoF, in terms of concurrence \cite{conc1, conc2}, has been obtained for the two-qubit system \cite{conc2}. A desirable property of an entanglement measure is its {\em additivity}. We say a quantity \(f\) is additive, if it satisfies the relation

\begin{equation}
f(\rho \otimes \sigma) = f(\rho) + f(\sigma),
\end{equation}

for product quantum states. For example, while squashed entanglement \cite{squashed-ent} is additive, negativity \cite{negativity} is not. We know that ${\cal N}(\rho \otimes \sigma) = {\cal N}(\rho)({\cal N}(\sigma) + 1) + {\cal N}(\sigma)({\cal N}(\rho) + 1)$ \cite{asu-activation}. EoF is trivially additive for pure product states. Further, it has been shown to be additive for some particular classes of states \cite{additivity-exmp}. Numerical calculations also support this claim. However, it is not known whether EoF satisfies additivity in general although there is a long-standing conjecture that EoF is additive, i.e., the EoF of the composite system equals the sum of the EoF's of its parts. If the additivity of EoF is true, then the entanglement cost of a state $\rho$ \cite{ent-cost}, $E_C(\rho) = \lim_{n \rightarrow \infty} E_F(\rho^{\otimes n})/n$, will be equal to the entanglement of formation. 
Let  $\rho_{XY}$ be the joint quantum state of a bipartite system $XY$. Then, another property called {\em strong superadditivity},
\begin{equation}
f(\rho_{XY}) \geq f(\rho_{X}) + f(\rho_{Y}),
\end{equation}
is quite interesting. 
The strong superadditivity of EoF implies additivity of EoF \cite{equivalence1, equivalence2}. 
It also implies the additivity of the Holevo-Schumacher-Westmorland classical capacity of a quantum channel \cite{equivalence2}.
In Ref. \cite{strong-super}, authors have shown the strong superadditivity of EoF assuming that additivity of EoF is true. That is, the additivity of EoF implies the strong superadditivity.
In quantum information theory, there are several open problems on the additivity of certain quantities \cite{additivity-prob}. It was conjectured \cite{equivalence1, equivalence2} that EoF satisfies strong superadditivity. Numerical analysis supports this fact. However, studies of Shor \cite{equivalence3-shor} and Hastings \cite{hastings} rule out this conjecture.  Shor showed that four additivity conjectures, namely, the conjectures of additivity of the minimum output entropy of a quantum channel, additivity of the Holevo expression for the
classical capacity of a quantum channel, additivity of the entanglement of formation, and
strong superadditivity of the entanglement of formation, are either all true or all false. 
Hastings in \cite{hastings} showed that all of these conjectures are false, by constructing a counterexample to the additivity conjecture for minimum output entropy. But, we haven't witnessed any violation of the strong superadditivity of EoF thus far. This motivated us to explore conditions under which EoF is strongly superadditive. In this paper we provide conditons for strong superadditivity of EoF, and identify classes of quantum states that are strongly superadditive.\\ 

{\it Notation}.-- 
Let $A_kB_k~(k=1,2)$ are two separate bipartite systems, and $\rho_{A_1B_1A_2B_2}$ be the joint quantum state of $A_1B_1$ and $A_2B_2$. In this paper, we will always consider entanglement of formation across the cut $A|B$.\\

{\it Conditions}.-- 
Below we give three conditions under which entanglement of formation is strongly superadditive.\\

\textbf{Condition 1.}
For a pure state $|\chi\rangle_{A_1B_1A_2B_2} \in \mathcal{H}_{A_1} \otimes \mathcal{H}_{B_1} \otimes \mathcal{H}_{A_2} \otimes \mathcal{H}_{B_2}$, entanglement of formation is strongly superadditive if the reduced density matrix, $\rho^{\chi}_{A_1A_2} = \mbox{tr}_{B_1B_2} |\chi\rangle_{A_1B_1A_2B_2} \langle \chi |$, admits one of the following forms:\\
\begin{align}
\rho^{\chi}_{A_1A_2} &= \sum_i p_i \mbox{tr}_{B_1} |\psi_i\rangle_{A_1B_1} \langle \psi_i | \otimes |i\rangle_{A_2} \langle i |, \\
\rho^{\chi}_{A_1A_2} &= \sum_i p_i  |i\rangle_{A_1} \langle i | \otimes \mbox{tr}_{B_2} |\phi_i\rangle_{A_2B_2} \langle \phi_i |.
\end{align}

\textit{Proof.} This follows systematically from the definition of EoF, $E_F(|\chi \rangle_{A_1A_2B_1B_2}) = S(\rho^{\chi}_{A_1A_2})$, the ``strong concavity'' of von Neumann entropy \cite{fan},  
\begin{align}
S(\sum_i p_i \rho^i_1 \otimes \rho^i_2) &\geq \sum_i p_i S(\rho^i_1) + S(\sum_i p_i \rho^i_2), \\
S(\sum_i p_i \rho^i_1 \otimes \rho^i_2) &\geq S(\sum_i p_i \rho^i_1) + \sum_i p_i S(\rho^i_2),
\end{align}
and the fact that for $\rho_{XY} = \sum_j q_j |\xi_j\rangle_{XY} \langle \xi_j |$, $E_F(\rho_{XY}) = \mbox{inf} \sum_j q_j S(\mbox{tr}_Y |\xi_j\rangle_{XY} \langle \xi_j |) \leq \sum_j q_j S(\mbox{tr}_Y |\xi_j\rangle_{XY} \langle \xi_j |) \leq S(\rho_X)$.
\hfill $\blacksquare$\\

\textbf{Condition 2.}
For a pure state $|\chi\rangle_{A_1B_1A_2B_2} \in \mathcal{H}_{A_1} \otimes \mathcal{H}_{B_1} \otimes \mathcal{H}_{A_2} \otimes \mathcal{H}_{B_2}$ with the Schmidt decomposition \cite{nielsen, preskill}, $|\chi\rangle_{A_1B_1A_2B_2} = \sum_i \sqrt{\lambda_i} |\psi_i\rangle_{A_1B_1} \otimes |\phi_i\rangle_{A_2B_2}$, entanglement of formation is strongly superadditive if $S(\rho^{\chi}_{A_1A_2}) \geq R(\chi)$, where $R(\chi) = \sum_i \lambda_i S(\mbox{tr}_{B_1} |\psi_i\rangle_{A_1B_1} \langle \psi_i | \otimes \mbox{tr}_{B_2} |\phi_i\rangle_{A_2B_2} \langle \phi_i |)$ (see \cite{note1}).

\textit{Proof.} 
This is because
\begin{eqnarray}
&&E_F(|\chi \rangle_{A_1A_2B_1B_2}) = S(\rho^{\chi}_{A_1A_2}) \nonumber \\
&\geq & \sum_i \lambda_i S(\mbox{tr}_{B_1} |\psi_i\rangle_{A_1B_1} \langle \psi_i | \otimes \mbox{tr}_{B_2} |\phi_i\rangle_{A_2B_2} \langle \phi_i |) \nonumber \\
&=&  \sum_i \lambda_i [S(\mbox{tr}_{B_1} |\psi_i\rangle_{A_1B_1} \langle \psi_i |) + S(\mbox{tr}_{B_2} |\phi_i\rangle_{A_2B_2} \langle \phi_i |)] \nonumber \\
&=&  \sum_i \lambda_i [E_F(|\psi_i\rangle_{A_1B_1}) + E_F(|\phi_i\rangle_{A_2B_2})] \nonumber \\
&\geq&  E_F(\rho^{\chi}_{A_1B_1}) + E_F(\rho^{\chi}_{A_2B_2}),
\label{eof-gen}
\end{eqnarray}
where the first inequality follows from the assumption, the second equality is due to the additivity of von Neumann entropy for product states, the third equality follows from the definition of EoF for a pure state, 
and the last inequality is due to the fact that $\rho^{\chi}_{A_1B_1} = \sum_{i} \lambda_i  |\psi_i \rangle_{A_1B_1} \langle \psi_i |$ and $\rho^{\chi}_{2} = \sum_{i} \lambda_i  |\phi_i \rangle_{A_2B_2} \langle \phi_i |$ may not be the optimal decompositions.
\hfill $\blacksquare$\\
Note that {\em Condition 2} is ``sufficient'' only because EoF may be strongly supperadditive even if it is violated. For example, for $|\chi\rangle_{A_1B_1A_2B_2} = \frac{1}{\sqrt{6}}(|00 \rangle_{A_1B_1} \otimes |01 \rangle_{A_2B_2} + |11 \rangle_{A_1B_1} \otimes |10 \rangle_{A_2B_2}) + \sqrt{\frac{2}{3}} |\psi^{+}\rangle_{A_1B_1} \otimes |\phi^{+}\rangle_{A_2B_2}$, where 
$|\phi^{\pm}\rangle = \frac{|00\rangle \pm |11\rangle}{\sqrt{2}}$ and $|\psi^{\pm}\rangle = \frac{|01\rangle \pm |10\rangle}{\sqrt{2}}$, we have $S(\rho^{\chi}_{A_1A_2}) = 1.25163$ and $R(\chi) = 4/3$ but $E_F(\rho^{\chi}_{A_1B_1}) + E_F(\rho^{\chi}_{A_2B_2}) = 0.374597$.\\

\textbf{Condition 3.}
Entanglement of formation is strongly superadditive for $\rho_{A_1B_1A_2B_2}$ if the reduced density matrices 
$\rho_{A_kB_k}~(k=1,2)$ are separable/diagonal.\\

\begin{center}
\begin{figure}[htb]
\subfloat[]{\includegraphics[width=2.7in, angle=0]{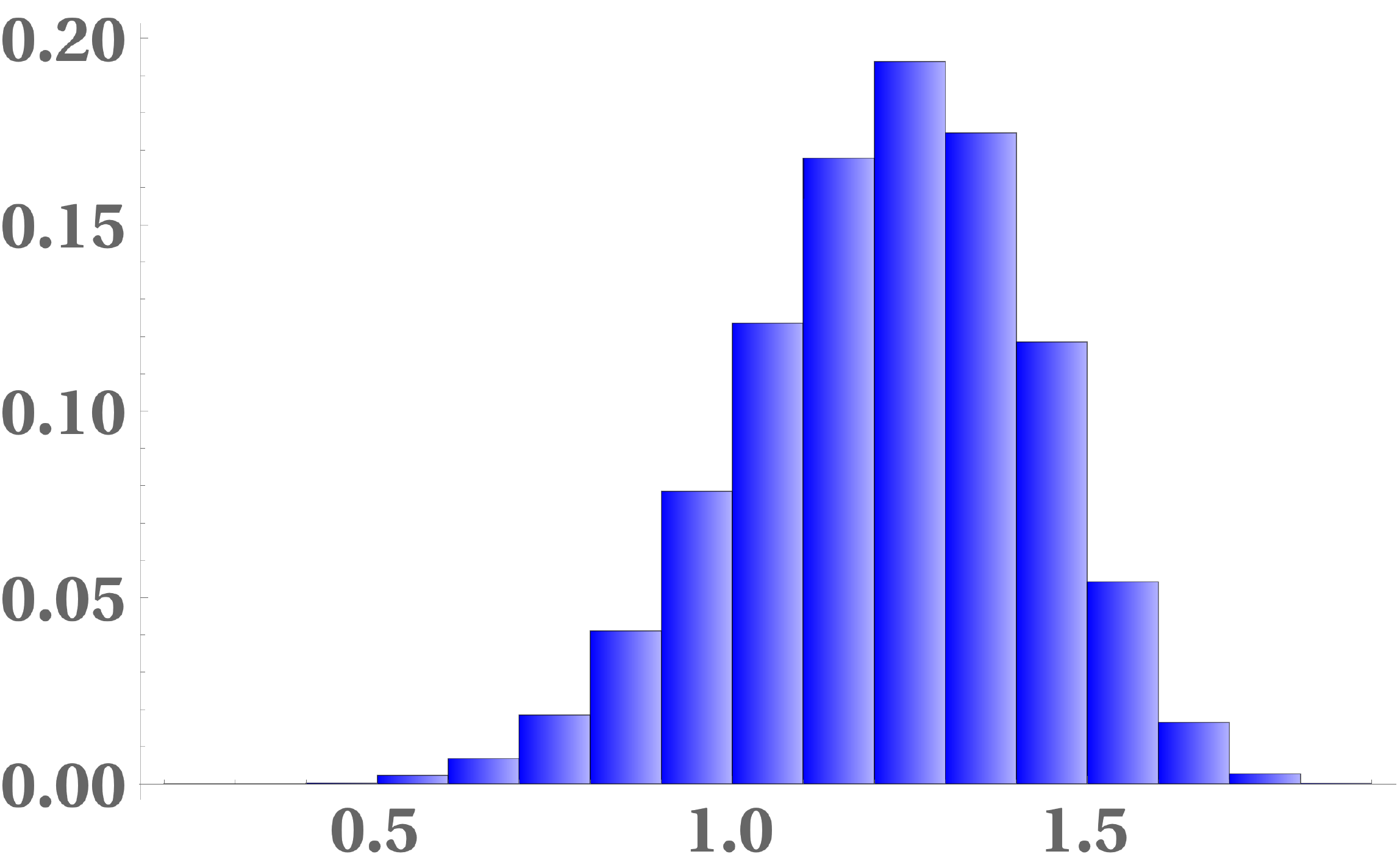}}\\
\subfloat[]{\includegraphics[width=2.7in, angle=0]{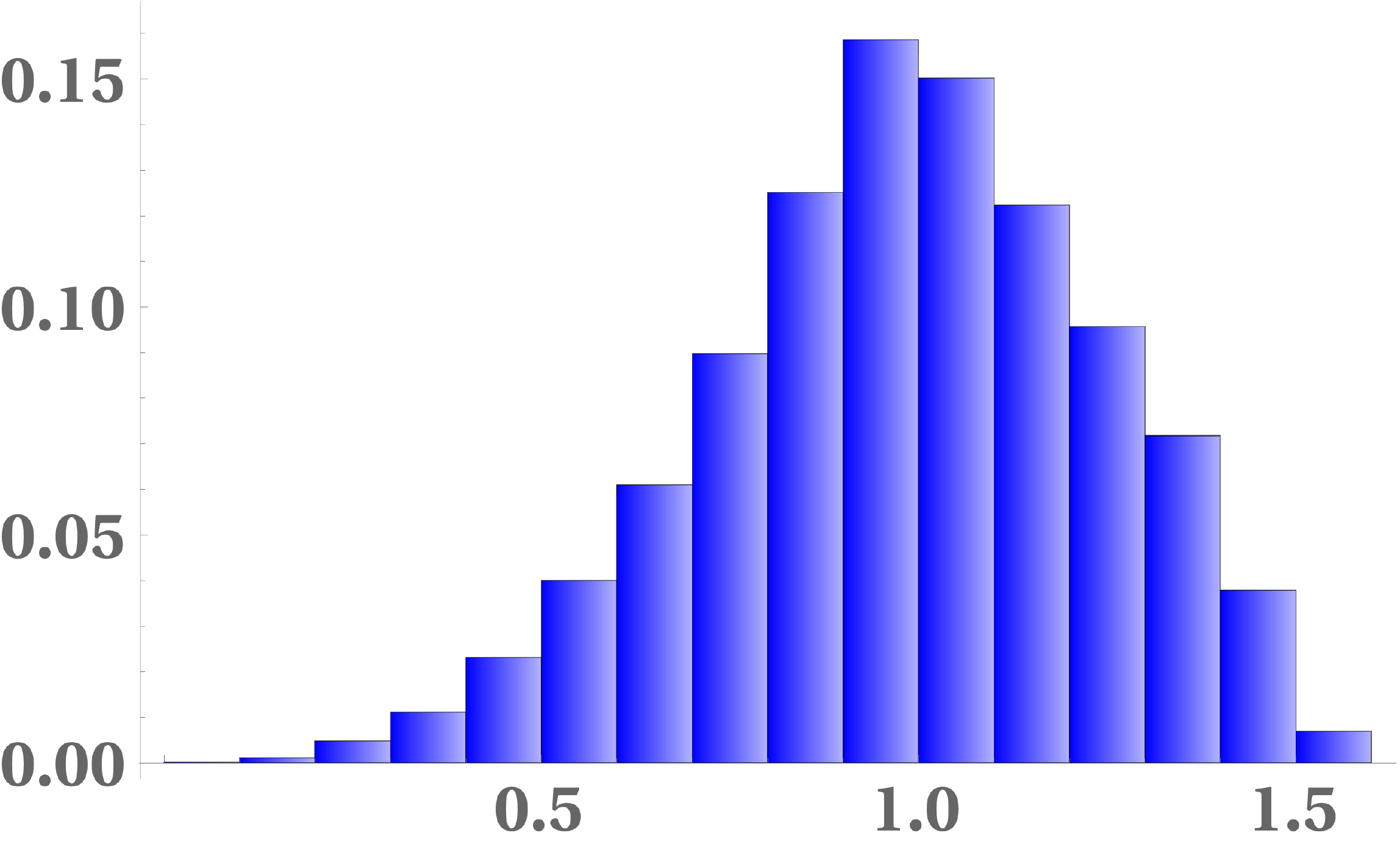}}
\caption{The strong superadditivity of entanglement of formation. We Haar uniformly generate $5 \times 10^4$ pure four-qubit (a) random, and (b) symmetric states, for generating histograms. Symmetric states are the linear combinations of Dicke states \cite{dicke}. The vertical axis represents the ``probability'' of occurance of a randomly generated four-qubit state in the corresponding range of $\Delta E_F = E_F(|\chi\rangle_{A_1B_1A_2B_2}) - E_F(\rho^{\chi}_{A_1B_1}) - E_F(\rho^{\chi}_{A_2B_2}) \geq 0$, on the horizontal axis. While the vertical axis is dimensionless, the horizontal axis is in ebits.}
\label{venn234scan}
\end{figure}
\end{center}

{\it Examples}.-- 
Based on above conditions, we identify classes of states for which EoF is strongly superadditive.\\
(i) Let $\{|i\rangle\}$ and $\{|j\rangle\}$ be orthonormal sets of vectors. Then, for $|\chi\rangle_{A_1B_1A_2B_2} = \sum_{ij} \sqrt{\lambda_{ij}} |ii\rangle_{A_1B_1} |jj\rangle_{A_2B_2}$,
\begin{align}
\rho^{\chi}_{A_1A_2} &= \sum_{ij} \lambda_{ij} |i\rangle_{A_1} \langle i | \otimes |j\rangle_{A_2} \langle j| \nonumber \\
&= \sum_{i} \lambda_{i} |i\rangle_{A_1} \langle i | \otimes \mbox{tr}_{B_2} |\phi_{i}\rangle_{A_2B_2} \langle \phi_{i}|, \nonumber
\end{align}
where $|\phi_{i}\rangle_{A_2B_2} = \frac{1}{\sqrt{\lambda_{i}}} \sum_j \sqrt{\lambda_{ij}} |jj\rangle_{A_2B_2}$, and $\sum_j \lambda_{ij} = \lambda_i$.\\
(ii) States $|\chi\rangle_{A_1B_1A_2B_2} = \sum_{ij} \sqrt{\lambda_{ij}} |ij\rangle_{A_1B_1} |ij\rangle_{A_2B_2}$, where $\{|i\rangle\}$ and $\{|j\rangle\}$ are orthonormal sets of vectors, satisfy both {\em Conditions 2 \& 3}.\\
(iii) Four-qubit X-states and four-qudit generalized GHZ states, $|\chi\rangle_{A_1B_1A_2B_2} = \sum_i \sqrt{\lambda_i} |iiii\rangle_{A_1B_1A_2B_2}$, satisfy {\em Condition 3}.\\

\textbf{Remark.} \textit{EoF is strongly superadditive for $\rho_{A_1B_1A_2B_2}$ if EoF is strongly superadditive for all the pure states in the optimal decomposition of $\rho_{A_1B_1A_2B_2}$.}\\

{\it Conclusion}.--
We have provided conditons for strong superadditivity of entanglement of formation, and identified some classes of states that are strongly superadditive. We also investigated numerically strong superadditivity of entanglement of formation for four-qubit pure states. \\
 


\begin{thebibliography}{99}
\bibitem{note1} Note that Schmidt decomposition is not unique. For example, $|\chi \rangle_{A_1B_1A_2B_2} = \frac{1}{\sqrt{2}}(|\phi^{+} \rangle_{A_1B_1} \otimes |\psi^{+} \rangle_{A_2B_2} + |\phi^{-} \rangle_{A_1B_1} \otimes |\psi^{-} \rangle_{A_2B_2}) = \frac{1}{\sqrt{2}}(|00 \rangle_{A_1B_1} \otimes |01 \rangle_{A_2B_2} + |11 \rangle_{A_1B_1} \otimes |10 \rangle_{A_2B_2})$. Here 
$|\phi^{\pm}\rangle = \frac{|00\rangle \pm |11\rangle}{\sqrt{2}}$ and $|\psi^{\pm}\rangle = \frac{|01\rangle \pm |10\rangle}{\sqrt{2}}$. While the former Schmidt decomposition violates {\em Condition 2}: $S(\rho^{\chi}_{A_1A_2}) = 1$ and $R(\chi) = 2$, the latter Schmidt decomposition satisfies {\em Condition 2}: $S(\rho^{\chi}_{A_1A_2}) = 1$ and $R(\chi) = 0$. Therefore, when the Schmidt decomposition is not unique, we will consider that decomposition for which $R(\chi)$ is minimum.
\end{thebibliography}
\end{document}